\documentclass[twocolumn,groupedaddress,showpacs,nofootinbib]{revtex4-1}

\usepackage[colorlinks,pdfstartview=FitH]{hyperref}
\hypersetup{linkcolor=blue,citecolor=blue,filecolor=black,urlcolor=blue}

\usepackage{amsmath,amssymb,bm}
\usepackage{graphicx}
\usepackage{amsfonts}
\usepackage{color}

\usepackage{amsfonts}
\usepackage{bbm}
\usepackage{latexsym}

\begin{document}

\baselineskip=15pt \parskip=5pt

\vspace*{3em}

\title{Dark photon portal dark matter with the 21-cm anomaly}

\author{Lian-Bao Jia }
\email{jialb@mail.nankai.edu.cn}
\affiliation{School of Science, Southwest University of Science and Technology, Mianyang
621010, China}

\begin{abstract}

A strong absorption profile was reported by the EDGES Collaboration, which indicates the hydrogen gas being colder than expected. It could be signatures of non-gravitational interactions between normal matter and dark matter (DM), and a potential explanation is that a small fraction of millicharged DM scatters with normal matter, with the DM mass in tens of MeV. To obtain the small fraction of millicharged DM and meanwhile being tolerant with by the constraints, the dark photon portal scalar and vector millicharged DM are explored in this paper. We consider that the mass of dark photon is slightly above twice of the millicharged DM mass, and thus the millicharged DM predominantly annihilates in p-wave during the freeze-out period, with the annihilation being enhanced near the resonance. The dark photon mainly decays into millicharged DM, and couplings of dark photon with SM particles could be allowed by the lepton collision experiments. The corresponding parameter spaces are derived. Future lepton collision experiments can be employed to search for millicharged DM via the production of the invisible dark photon.

\end{abstract}

\maketitle

\section{Introduction}

In the Universe, about 84\% of the cosmological matter density is contributed by dark matter (DM) \cite{Ade:2015xua}, while little has been known about particle properties of DM by now. Recently, an absorption profile around 78 MHz in the sky-averaged spectrum was reported by the EDGES Collaboration \cite{Bowman:2018yin}, and the magnitude of the absorption was enhanced with 3.8$\sigma$ discrepancy. This enhancement would indicate the hydrogen gas being colder than expected, and it may be signatures of non-gravitational interactions between normal matter and DM from the cosmic dawn \cite{Bowman:2018yin,Barkana:2018lgd,Fialkov:2018xre,Mirocha:2018cih,Li:2018kzs}.

To cool the hydrogen gas via the scatterings between DM and hydrogen, residual electrons and protons, the mass of DM should be not much heavier than the hydrogen mass.\footnote{For other scenarios, such as modifying the radio background, see e.g., Refs. \cite{Feng:2018rje,Fraser:2018acy,Pospelov:2018kdh}.} Meanwhile, to explain the absorption profile, velocity-independent scatterings seem to be in tension with the cosmic microwave background (CMB) observation, and velocity-dependent scatterings are available \cite{Barkana:2018lgd,Fialkov:2018xre}, with the cross sections $\propto v^{-4}$. Possible new light mediators with masses $\lesssim$ 10$^{-3}$ eV are not favored by constraints from CMB and the big bang nucleosynthesis \cite{Ade:2015xua,Kawasaki:1992kg,Foot:2014uba,Berlin:2018sjs,Barkana:2018qrx}, and a feasible scenario is that a small fraction (about 0.003$-$0.02) of DM is millicharged (DM particle carries an electric charge $\eta e$ with $\eta \sim 10^{-6} - 10^{-4}$) \cite{Munoz:2018pzp,Berlin:2018sjs,Barkana:2018qrx,Mahdawi:2018euy}, with DM mass about 10$-$80 MeV and photon as the mediator in the scattering of cooling the hydrogen. For the vector-vector current scattering mediated by photon (in the nonrelativistic limit), the parameter spaces are nearly the same not only for scalar and fermionic millicharged DM \cite{Berlin:2018sjs}, but also for vectorial millicharged DM (see, e.g., Ref. \cite{Jia:2013lza}).

To avoid the overproduction of DM in the early universe, new annihilation mechanisms are needed to obtain the required DM relic abundance. Besides DM being millicharged, here we consider that DM is also dark charged, which can have other couplings to the standard model (SM) sector, e.g., via the photon-dark photon kinetic mixing $ \frac{1}{2} (\epsilon / \cos \theta_W) F_{\mu\nu} A'^{\mu\nu}$, where $A'$ is the dark photon field (see e.g., Refs. \cite{Okun:1982xi,Galison:1983pa,Holdom:1985ag,Fayet:1990wx,Foot:2014osa,Bilmis:2015lja,Feng:2015hja,Chun:2009zx} for more). In addition, the CMB observation \cite{Ade:2015xua,Slatyer:2015jla} and the 21-cm absorption profile from the cosmic dawn \cite{Liu:2018uzy,DAmico:2018sxd} set stringent constraints on s-wave annihilations of DM with masses in the MeV scale. A possible way is that the millicharged DM predominantly annihilates in p-wave during the freeze-out period, and thus the scalar and vector DM with masses being lighter than the dark photon are of our concern.

Due to a small fraction of DM being millicharged, the couplings of dark photon with SM particles may be not very small, and this will be restricted by the lepton collision experiments \cite{Izaguirre:2013uxa,Banerjee:2016tad,Banerjee:2017hhz,Lees:2017lec}. Here we consider the case that the mass of dark photon is slightly above twice of the millicharged DM mass. In this case, dark photon can mainly decay into millicharged DM. The annihilations of millicharged DM are significantly enhanced near the resonance, and the couplings of dark photon with SM particles could be allowed by the lepton collision experiments. The corresponding parameter space will be derived in this paper.

\section{Interactions and transitions}

Here we consider the photon and dark photon $A'$ mediate the transitions between millicharged DM and SM sector. The interaction of $A'$ boson with SM charged fermion is taken as
\begin{eqnarray}
\mathcal{L}_i^{SM} = e \epsilon  A'_{\mu} J_{\mathrm{em}}^{\mu}    .
\end{eqnarray}
For scalar (vector) millicharged DM $\phi$ ($V$), the electric charge is taken as $\eta e$ ($\eta \sim$ $10^{-6}-10^{-4}$ \cite{Munoz:2018pzp,Berlin:2018sjs,Barkana:2018qrx}), and the dark charge is $e_D^{}$. Here we focus on the case that the main decay products of $A'$ are invisible, i.e., $A'$ mainly decaying into DM pairs $\phi \phi^\ast$ ($V V^\ast$) with the mass $2 m_\phi (2 m_V) < m_{ A'}$. To enhance DM annihilations during millicharged DM freeze-out, here we consider that $m_{ A'}$ is slightly above $2 m_\phi$ $(2 m_V) $. Thus, the annihilations of DM can be significantly enhanced close to the resonance, and the required large annihilation cross section is mainly contributed by $A'$. For the DM mass range of concern, the p-wave annihilation process $\phi \phi^\ast$ ($V V^\ast$) $\to A' $ $\to e^+ e^-$ is predominant during DM freeze-out.

Now we formulate the millicharged DM annihilations mediated by $A'$. For scalar millicharged DM $\phi$, the annihilation cross section of the process $\phi \phi^\ast$ $\to A' $ $\to e^+ e^-$ is
\begin{eqnarray}
\sigma_{\mathrm{ann}} v_r = \frac{1}{2} \frac{e_D^2 e^2 \epsilon^2 }{12 \pi (s - 2 m_\phi^2)} \frac{s (s - 4 m_\phi^2 )}{(s -  m_{A'}^2)^2 + m_{A'}^2 \Gamma_{A'}^2},
\end{eqnarray}
where $v_r$ is the relative velocity of the annihilating DM pair, and the factor $\frac{1}{2}$ is included here for the required $\phi \phi^\ast$ pair in DM annihilations. $s$ is the total invariant mass squared, with $s = $ $4 m_\phi^2 + m_\phi^2 v_r^2 + \mathcal{O} (v_r^4)$ in the nonrelativistic limit. Here the electron mass is negligible for the DM mass of concern. $\Gamma_{A'}$ is the decay width of $A'$, with
\begin{eqnarray}
\Gamma_{A'} \approx \frac{e_D^2}{48 \pi} m_{A'} (1 - \frac{4 m_\phi^2}{m_{A'}^2 })^\frac{3}{2}  .
\end{eqnarray}

For vector millicharged DM $V$, the annihilation cross section of the process $V V^\ast$ $\to A' $ $\to e^+ e^-$ is
\begin{eqnarray}
\sigma_{\mathrm{ann}} v_r \simeq \frac{1}{2} \frac{e_D^2 e^2 \epsilon^2 }{54 \pi (s - 2 m_V^2)} \frac{13 s (s - 4 m_V^2 )}{(s -  m_{A'}^2)^2 + m_{A'}^2 \Gamma_{A'}^2} .
\end{eqnarray}
In the nonrelativistic limit, one has $s = $ $4 m_V^2 + m_V^2 v_r^2 + \mathcal{O} (v_r^4)$. The corresponding $\Gamma_{A'}$ is
\begin{eqnarray}
\Gamma_{A'} \approx \frac{e_D^2}{48 \pi} m_{A'} (1 - \frac{4 m_V^2}{m_{A'}^2 })^\frac{3}{2} (3 + \frac{5 m_{A'}^2}{m_V^2} + \frac{m_{A'}^4}{4 m_V^4}) .
\end{eqnarray}

Here we give a brief discussion about the photon mediated transitions. The annihilation mode $\phi \phi^\ast$ ($V V^\ast$) $\to \gamma $ $\to e^+ e^-$ is a p-wave process, which is suppressed by the millicharge parameter $\eta^2$, and the corresponding thermally averaged annihilation cross section is in a range about $10^{-32} - 10^{-26}$ cm$^3$/s (much smaller than the value required by the small fraction of millicharged DM). The annihilation mode $\phi \phi^\ast$ ($V V^\ast$) $\to \gamma \gamma$ is a s-wave process, and this process is deeply suppressed by $\eta^4$. Thus, we can neglect the photon's contribution during millicharged DM freeze-out. For the millicharged DM of concern, the corresponding energy injection from the process $\phi \phi^\ast$ ($V V^\ast$) $\to \gamma \gamma$ is allowed by the CMB observation \cite{Ade:2015xua,Slatyer:2015jla} and the 21-cm absorption profile \cite{Bowman:2018yin,Liu:2018uzy}.

\section{Numerical analysis}

As given by the Introduction, to cool the hydrogen gas indicated by the EDGES observation, the fraction of millicharged DM $f_\mathrm{DM}$ is about 0.003$-$0.02, with DM mass $\sim$ 10$-$80 MeV. For teens MeV DM, the annihilation products of DM could heat the electron-photon plasma in the early universe after the electron neutrino decoupling, and this would reduce the effective number of the relativistic neutrinos $N_{\mathrm{eff}}$. Considering the recent Planck observations on $N_{\mathrm{eff}}$ \cite{Ade:2015xua}, one has the relation of the DM mass \cite{Jia:2016uxs}: $m_\phi \gtrsim$ 10.4 MeV for scalar DM, and $m_V \gtrsim$ 13.6 MeV for vector DM. Thus, the following mass range of the millicharged DM is adopted here: 10.4 $ \lesssim m_\phi \lesssim$ 80 MeV for scalar DM, and 13.6 $ \lesssim m_V \lesssim$ 80 MeV for vector DM.

\begin{figure}[htbp!]
\includegraphics[width=0.4\textwidth]{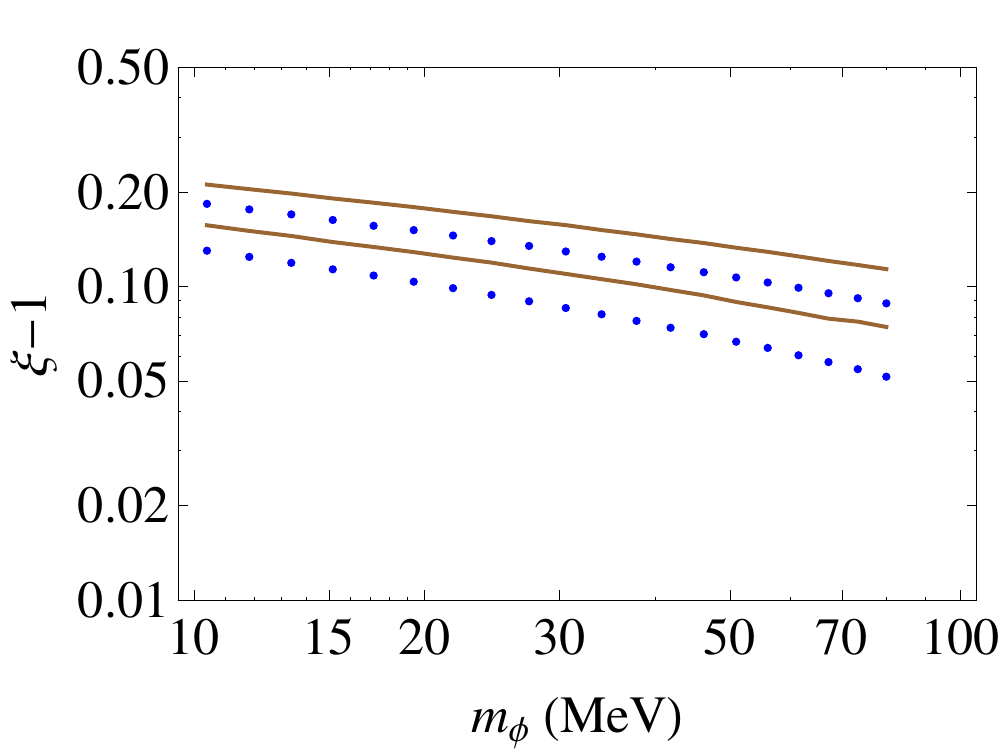} \vspace*{-1ex}
\includegraphics[width=0.4\textwidth]{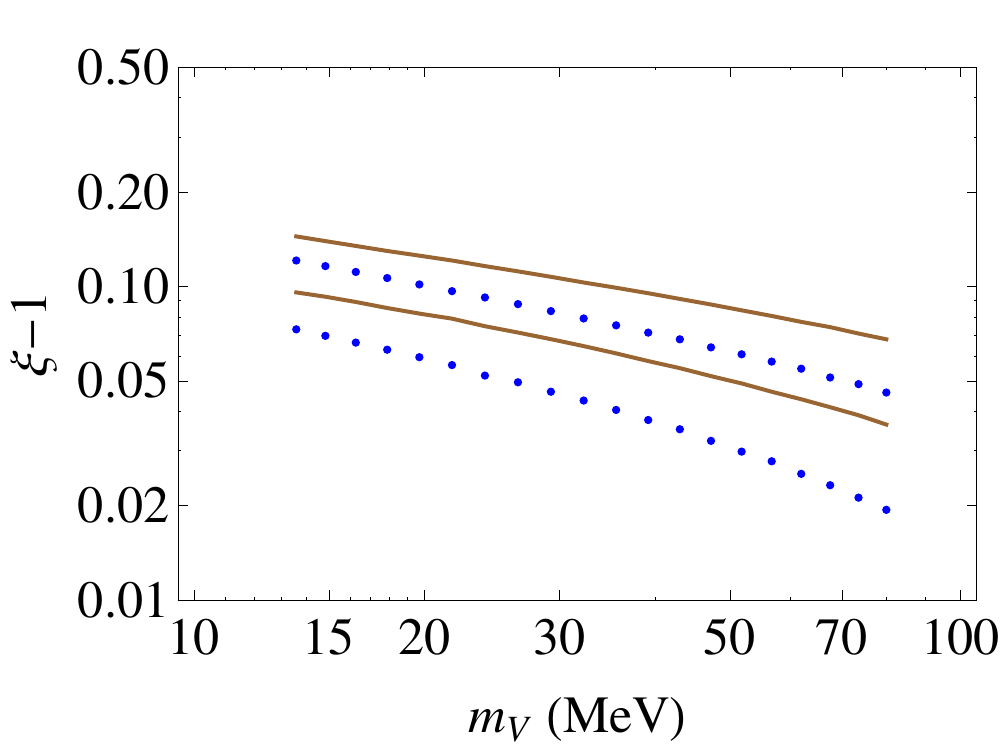} \vspace*{-1ex}
\caption{The required value of $\xi - 1$ for scalar and vector millicharged DM with the EDGES observation. The coupling parameter $e_D^{} \epsilon$ $= 10^{-4}$ is taken here. The mass range of the scalar millicharged DM is 10.4 $ \lesssim m_\phi \lesssim$ 80 MeV, and that is 13.6 $ \lesssim m_V \lesssim$ 80 MeV for vector millicharged DM. The dotted, solid curves are for the case of $e_D^{} =$ 1, 0.5 respectively. In each type curves (dotted, solid curves), the upper one, lower one are corresponding to $f_\mathrm{DM} =$ 0.02, 0.003 respectively.}\label{dm-xi}
\end{figure}

The total relic density of DM is $\Omega_D h^2 =$ 0.1197 $\pm$ 0.0042 \cite{Ade:2015xua}. For scalar (vector) millicharged DM, to obtain the large annihilation cross section at the freeze-out epoch indicated by the small $f_\mathrm{DM}$, we consider the annihilation being close to the resonance.

Note $\xi = m_{ A'} / 2 m_\phi$ ($ m_{ A'} / 2 m_V$), and here $\xi$ is slightly above 1. For a given coupling parameter $e_D^2 \epsilon^2$, the annihilation cross section of millicharged DM will be increased with the reducing of $\xi$'s value (here the value $\xi -1 > 0$). With the perturbative requirement, one has $e_D^2 < 4 \pi$. In addition, the recent lepton collision experiments, such as BaBar \cite{Lees:2017lec} and NA64 \cite{Banerjee:2017hhz}, set an upper limit on $\epsilon$ in the search of the dark photon's invisible decay, i.e., $\epsilon \lesssim$ ($10^{-4} - 10^{-3}$) for $ m_{ A'}$ in a range of 20$-$200 MeV. Thus, a typical upper limit $e_D^{} \epsilon$ $\lesssim 10^{-4}$ is adopted here. In this case, to obtain the fraction $f_\mathrm{DM}$ of millicharged DM indicated by the EDGES experiment, the corresponding $\xi$ required is shown in Fig. \ref{dm-xi} (see Appendix \ref{appendix:dm-ann} for the calculation), with $e_D^{} \epsilon$ $= 10^{-4}$ for scalar and vector millicharged DM. The dotted, solid curves are for the value $e_D^{} =$ 1, 0.5 being adopted respectively, and the upper one, lower one in each type curves are corresponding to $f_\mathrm{DM} =$ 0.02, 0.003 respectively. For a given $e_D^2 \epsilon^2$, a smaller fraction $f_\mathrm{DM}$ is corresponding to a smaller $\xi -1$ value.

\begin{figure*}[htbp!]
\includegraphics[width=0.4\textwidth]{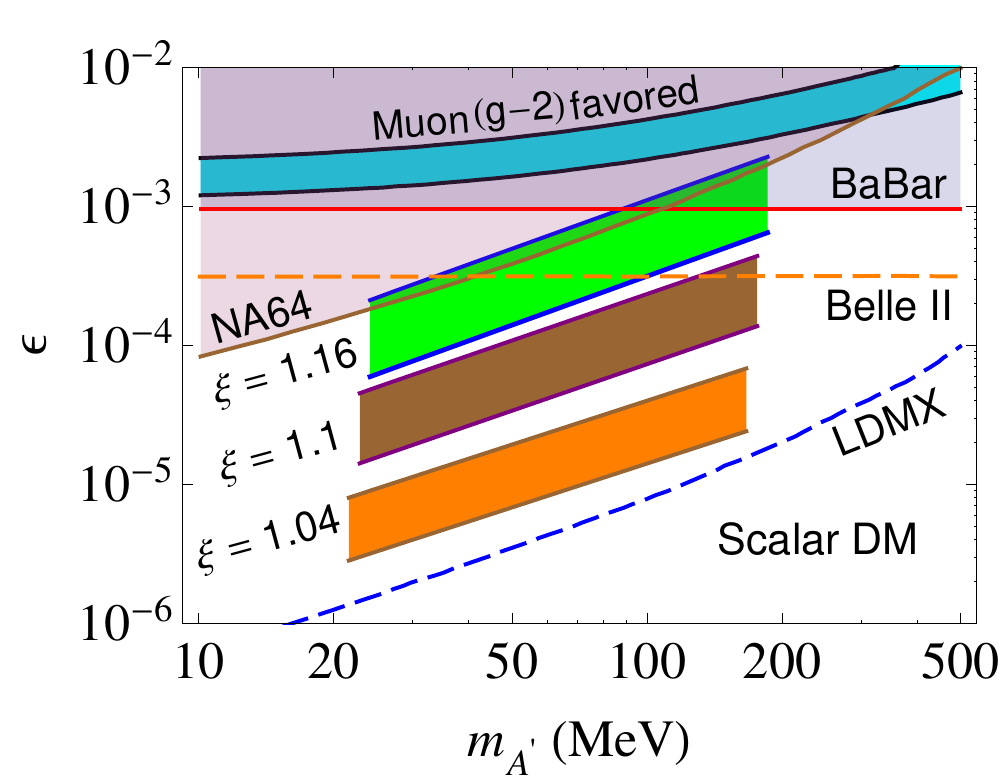} \vspace*{-1ex}
\includegraphics[width=0.4\textwidth]{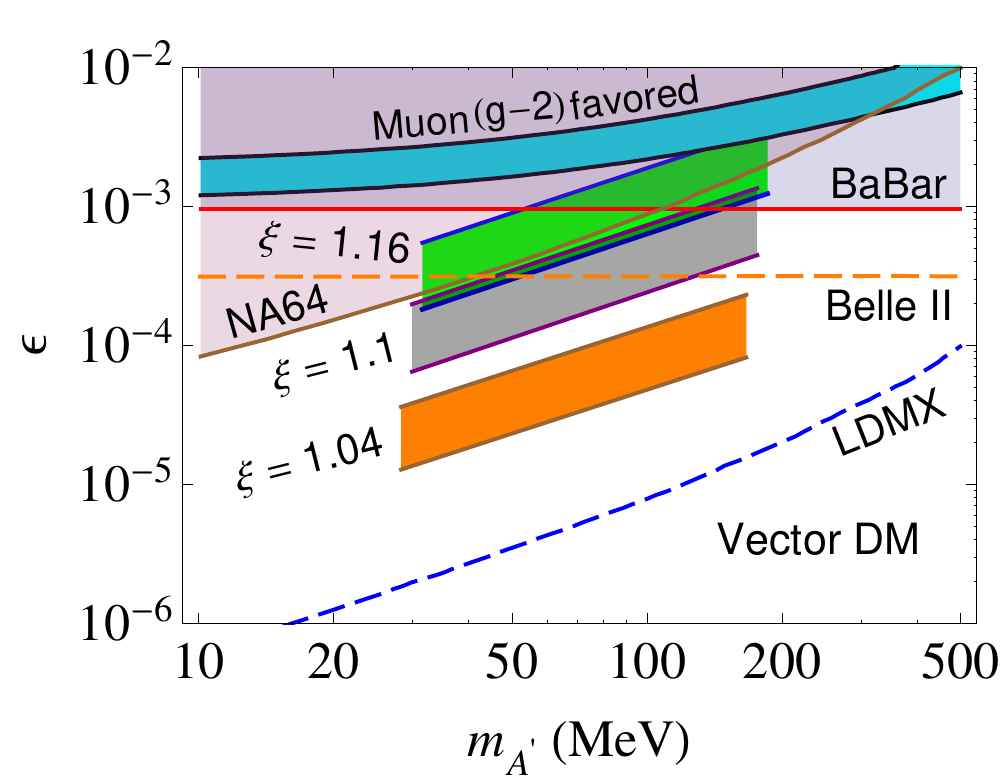} \vspace*{-1ex}
\caption{The value of $\epsilon$ as a function of $ m_{ A'}$ for scalar and vector millicharged DM with $\xi =$ 1.16, 1.1 and 1.04. Here $e_D^{} =$ 1 is adopted. For the band of a given $\xi$, the upper, lower limits are for $f_\mathrm{DM} =$ 0.003, 0.02 respectively, with the range of $f_\mathrm{DM}$ indicated by the 21-cm anomaly. The upper limit of $\epsilon$ with constraints from NA64 \cite{Banerjee:2017hhz} and BaBar \cite{Lees:2017lec}, and the region preferred by the muon g$-$2 \cite{Bennett:2006fi} are denoted in the figure. The upper and lower dashed curves are the expected sensitivity of Belle II 20 fb$^{-1}$ \cite{Park:2018bfb} and the ultimate reach of Light Dark Matter eXperiment (LDMX) \cite{Akesson:2018vlm}, respectively.}\label{dm-m}
\end{figure*}

The invisible dark photon (mainly decaying into scalar, vector millicharged DM) may be produced at the lepton collision experiments \cite{Izaguirre:2013uxa,Banerjee:2016tad,Banerjee:2017hhz,Lees:2017lec}, or the value of the kinetic mixing parameter $\epsilon$ would be restricted by the experiments. For a given $\xi$, to obtain the fraction of millicharged DM indicated by the 21-cm absorption, the corresponding range of $\epsilon$ is derived, as shown in Fig. \ref{dm-m}, with $\xi =$ 1.16, 1.1 and 1.04, and $e_D^{} =$ 1. For a given $\xi$, the upper, lower limits of each band are for $f_\mathrm{DM} =$ 0.003, 0.02 respectively. The recent results of NA64 \cite{Banerjee:2017hhz} and BaBar \cite{Lees:2017lec} set an upper limit on $\epsilon$. It can be seen that for $\xi =$ 1.16, $f_\mathrm{DM}$ in a range about 0.003$-$0.02 is allowed for scalar millicharged DM, while the parameter space for a small fraction of $f_\mathrm{DM}$ is excluded and $f_\mathrm{DM}$ close to 0.02 is left for vector millicharged DM. The lepton collision experiment does well in the search of $A'$ for the case of a small $f_\mathrm{DM}$ and a large $\xi$. The parameter spaces can be tested at future experiments, such as Belle II \cite{Park:2018bfb} and the Light Dark Matter eXperiment (LDMX) \cite{Akesson:2018vlm}.

Here we give a brief discussion about the detection of millicharged DM by underground experiments.
For tens MeV DM, the results of XENON10 \cite{Essig:2012yx,Essig:2017kqs} and COHERENT \cite{Ge:2017mcq} seem to be sensitive for the scattering mediated by massless mediators (or the mediator's mass being very tiny). However, for the millicharged DM of concern, the exclusion region is feasible for the millicharge parameter $\eta \lesssim 10^{-7}$ \cite{Barkana:2018qrx,Emken:2017erx}, accounting for the terrestrial effect when a charged particle penetrating the earth. Moreover, due to the magnetic fields in the Milky Way, the millicharged DM is expected to be evacuated from the Galactic disk \cite{Barkana:2018lgd,Chuzhoy:2008zy,McDermott:2010pa}, and hence will be absent from DM direct detections. Thus, the millicharged DM of concern can be allowed by the direct detections (see e.g., Ref. \cite{Barkana:2018qrx} for more).

\section{Conclusion and discussion}

The dark photon portal scalar/vector millicharged DM were studied in this paper, which could cool the gas and produce the 21-cm anomaly via photon mediated scattering. To obtain the small fraction $f_\mathrm{DM}$ of millicharged DM, we consider that the annihilations are mainly mediated by the dark photon during millicharged DM freeze-out, with the annihilations near the resonance and the parameter $\xi$ being slightly above 1. The annihilation mediated by the dark photon is a p-wave process, and thus it is tolerated by constraints from CMB and the 21-cm absorption profile.

The value of $\xi$ was derived for the required fraction $f_\mathrm{DM} =$ 0.003$-$0.02, with 10.4 $ \lesssim m_\phi \lesssim$ 80 MeV for scalar millicharged DM, and 13.6 $ \lesssim m_V \lesssim$ 80 MeV for vector millicharged DM. For a given coupling parameter $e_D^2 \epsilon^2$, a smaller fraction $f_\mathrm{DM}$ requires a smaller $\xi -1$ value. For a given $\xi$, the range of $\epsilon$ was derived for the $f_\mathrm{DM}$ of millicharged DM indicated by the 21-cm absorption. The DM with the millicharge parameter $\eta$ of concern could be absent in direct detection experiments. The lepton collision experiment does well in the search of millicharged DM via the production of the invisible dark photon, such as Belle II and LDMX, especially for the case of a small $f_\mathrm{DM}$ and a large $\xi$. We look forward to the search of millicharged DM at future lepton collision experiments.

\acknowledgments \vspace*{-3ex} This work was supported by National Natural Science Foundation of China  under the contract No. 11505144, and the Longshan Academic Talent Research Supporting Program of SWUST
under Contract No. 17LZX323.

\appendix

\section{$f_\mathrm{DM}$ and the parameter $\xi$}
\label{appendix:dm-ann}

The relic density of millicharged DM is $f_\mathrm{DM} \Omega_D$. For thermal freeze-out millicharged DM at temperature $T_f$, $f_\mathrm{DM}$ can be set by the thermally averaged annihilation cross section $\langle \sigma_{\mathrm{ann}} v_r \rangle$ via the relation \cite{Griest:1990kh}
\begin{eqnarray}
f_\mathrm{DM} \Omega_D h^2  \simeq  \frac{1.07 \times 10^9  ~ \mathrm{GeV}^{-1}}{J_{\mathrm{ann}} m_{\mathrm{Pl}} \sqrt{g_\ast}}  ,
\end{eqnarray}
with
\begin{eqnarray}
J_{\mathrm{ann}} =  \int^\infty_{x_f} \frac{\langle \sigma_{\mathrm{ann}} v_r \rangle}{x^2} \mathrm{d} x ,
\end{eqnarray}
and the parameter $x_f$ is $x_f = m_\phi / T_f$ $( m_V / T_f)$. Substituting $m_{ A'}  = 2 m_\phi \xi$ ($ 2 m_V \xi$) into $\langle \sigma_{\mathrm{ann}} v_r \rangle$, the relation between $\xi$ and $f_\mathrm{DM}$ can be obtained.

\end{document}